\documentclass[acmsmall]{acmart}

\AtBeginDocument{%
  \providecommand\BibTeX{{%
    \normalfont B\kern-0.5em{\scshape i\kern-0.25em b}\kern-0.8em\TeX}}}

\setcopyright{rightsretained}
\acmJournal{PACMHCI}
\acmYear{2021} \acmVolume{5} \acmNumber{CSCW2} \acmArticle{462} \acmMonth{10} \acmPrice{}
\acmDOI{10.1145/3479606}

\acmSubmissionID{1750}

\begin{document}

\title{Why Do People Participate in Small Online Communities?}

\author{Sohyeon Hwang}
\email{sohyeonhwang@u.northwestern.edu}
\orcid{0000-0001-8415-7395}
\affiliation{%
  \institution{Northwestern University}
  \city{Evanston}
  \state{Illinois}
  \country{USA}
}

\author{Jeremy Foote}
\email{jdfoote@purdue.edu}
\orcid{0000-0002-0078-2925}
\affiliation{%
  \institution{Purdue University}
  \city{West Lafayette}
  \state{Indiana}
  \country{USA}
}

\renewcommand{\shortauthors}{Hwang and Foote}

\begin{abstract}
  Many benefits of online communities---such as obtaining new information, opportunities, and social connections---increase with size. Thus, a ``successful'' online community often evokes an image of hundreds of thousands of users, and practitioners and researchers alike have sought to devise methods to achieve growth and thereby, success. On the other hand, small online communities exist in droves and many persist in their smallness over time. Turning to the highly popular discussion website Reddit, which is made up of hundreds of thousands of communities, we conducted a qualitative interview study examining how and why people participate in these persistently small communities, in order to understand why these communities exist when popular approaches would assume them to be failures. Drawing from twenty interviews, this paper makes several contributions: we describe how small communities provide unique informational and interactional spaces for participants, who are drawn by the hyperspecific aspects of the community; we find that small communities do not promote strong dyadic interpersonal relationships but rather promote group-based identity; and we highlight how participation in small communities is part of a broader, ongoing strategy to curate participants' online experience. We argue that online communities can be seen as nested niches: parts of an embedded, complex, symbiotic socio-informational ecosystem. We suggest ways that social computing research could benefit from more deliberate considerations of interdependence between diverse scales of online community sizes.
\end{abstract}



\maketitle

\section{Introduction}
\begin{figure}
    \centering
    \includegraphics[width=.8\textwidth]{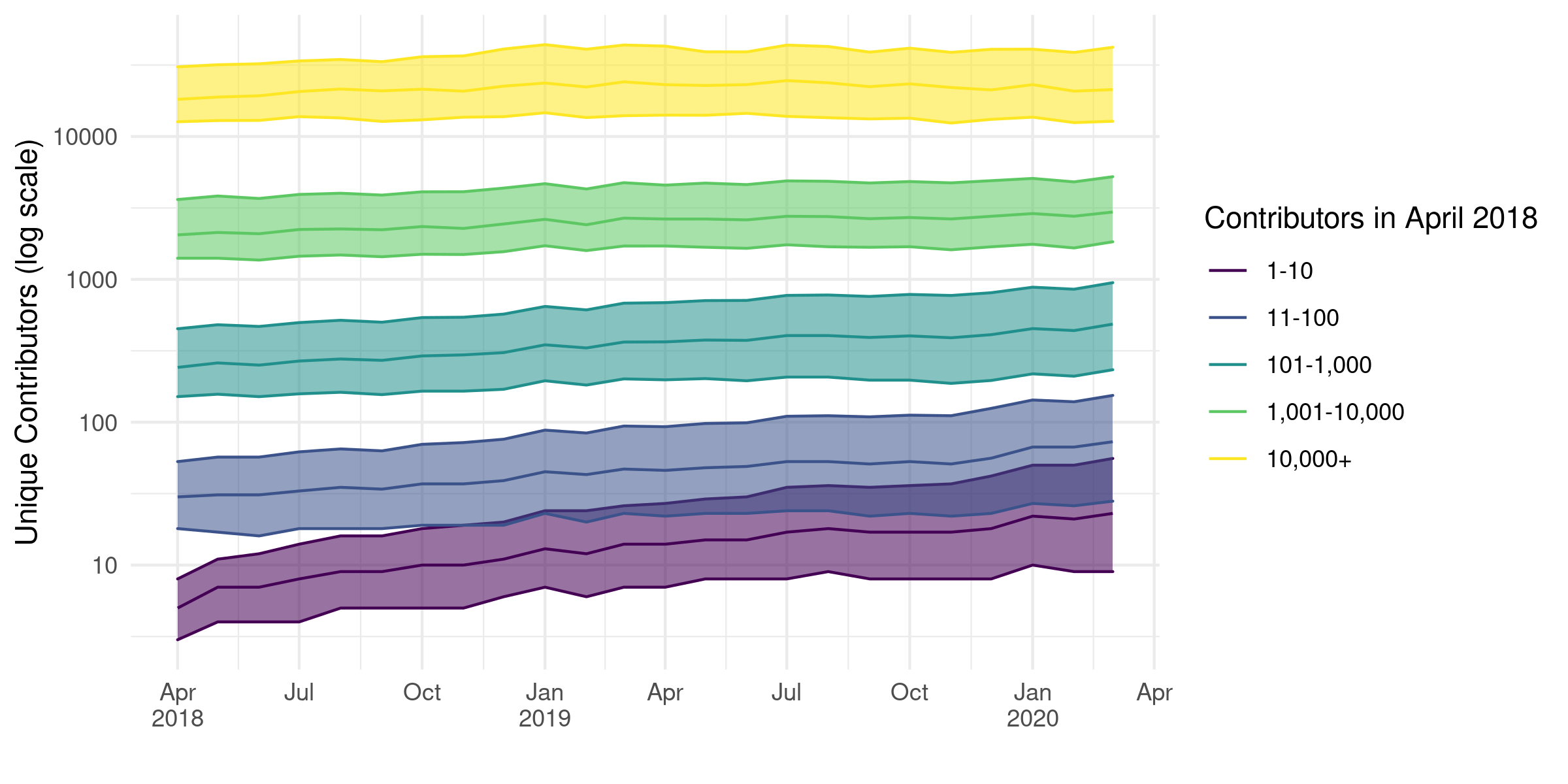}
    \caption{Size of consistently active subreddits over time (i.e., those with at least one comment per month from April 2018 to March 2020). Subreddits are grouped by their size in April 2018. Lines represent the median size each month, and ribbons show the first and third quartiles.}
    \label{fig:over_time}
\end{figure}

Size and growth are often natural indicators of whether an online community is sustainable and successful.
Many of the benefits that people seek from online communities---such as information, entertainment, or novelty---seem to increase with size. Works aimed at practitioners and researchers often use metaphors like a critical mass or network effects \cite[e.g.,][]{kraut_building_2012,butler_attractionselectionattrition_2014,butler_membership_2001} and assume that all communities seek growth, or at least that they would be better able to meet the needs of users if they grew larger.
Despite this, small online communities still exist in droves. On Reddit, one of the most popular online community sites, small communities not only exist in great numbers but many persist in their smallness over time (Figure \ref{fig:over_time}). These persistently small communities challenge the assumptions underlying many online community design recommendations for how to recruit more participants, encourage more contributions, and retain more users: that growth is imperative and that small communities will either become large or die off.

Why do so many communities remain small throughout their ``lives''? What value do participants get from small communities and how do these benefits relate to community size? Our understanding of online communities would be enriched by a deeper understanding of how motivations, participation dynamics, and perceptions vary across communities of different sizes. Rather than growth being an unalloyed good, we suggest that some benefits can be obtained \emph{only} through small communities and that users may intentionally seek out the kinds of experiences that small communities distinctly provide. 

In this study, we focus on the long tail of small communities, exploring how and why Reddit users participate in them, through a qualitative interview study. Drawing from conversations with twenty participants of various small online communities, this paper makes multiple contributions. First, we supply evidence that small communities provide qualitatively distinct benefits to their participants, such as expertise, trust, and a supportive community. Second, we find that small communities in our sample often provide participants with a sense of camaraderie or group identity but rarely serve as a source of dyadic relationships, contrary to what prior work would suggest. Third, we present a theory that small communities are enabled by and enable a robust ecosystem of semi-overlapping topical communities of different sizes and specificity. We end by reflecting on the role of size in online communities and the importance of examining and understanding communities across different scales of sizes.

\section{Background}

In the following sections, we situate this study in the broader social computing literature studying why people participate in online communities. We reflect on the motivations for participation and the benefits that people get from their participation, and then consider how these benefits could relate to the size of an online community, outlining why we would expect larger communities to have advantages over small ones. We end with a discussion of why small communities may exist regardless, noting a dearth of work in social computing explicitly focused on the unique advantages of small communities.

\subsection{Why people participate in online communities}
The term ``online community'' covers a broad range of activity in the social computing literature, and many different types of communities have been studied. As a result, CSCW literature has used the term to describe entire platforms such as Slashdot \cite{lampe_follow_2005}, a social news and discussion site, as well as distinct groups within those platforms such as Usenet newsgroups \cite{arguello_talk_2006}, Facebook groups \cite{lopez_consequences_2013}, and Reddit subreddit communities \cite{kiene_surviving_2016,lin_better_2017}. Unsurprisingly, because online communities are so diverse in their scopes, purposes, and topics, the motivations for participating in them are similarly wide-ranging. In this section, we give a brief overview of work seeking to understand participant motivations, paying particular attention to research on discussion-based and forum-style communities, which are the locus of our study.

Most of this research can be classified as taking either the \emph{user perspective} or the \emph{community manager perspective}. The \emph{user perspective} is represented by a major line of work rooted in social psychology, focusing on the motivations of individuals who take part in communities. This research has examined what motivates individuals to join, contribute to, and stay in online communities \cite{ling_using_2005,kraut_building_2012}. Scholars have noted the role of \emph{social identity} in fostering participation in virtual spaces for social groups or interests core to users' identities \cite{ammari_understanding_2015, li_slacktivists_2018, lingel_city_2014, zhang_community_2017};  
\emph{interpersonal interactions} in making participants feel welcome and heard \cite{backstrom_preferential_2008}; and \emph{personal and/or social bonds} in drawing individuals to communities \cite{ren_building_2012}. Other motivations include \emph{information-seeking and information-sharing} \cite{lu_encouraging_2011,jin_why_2015}, \emph{entertainment}, and \emph{self-discovery} \cite{lampe_motivations_2010}.

Some of this work on participant motivations has applied \textit{uses and gratifications theory} to help frame the varied, complex reasons individuals engage with different media and online communities \cite{dholakia_social_2004,lampe_motivations_2010}. Uses and gratifications theory posits that users will intentionally seek specific types of media to satisfy specific needs \cite{ruggiero_uses_2000,lampe_motivations_2010}. Under this theory, users may engage in different online communities for different reasons. This helps to explain the various and at times conflicting participant motivations that empirical studies have identified \cite{malinen_understanding_2015} and acknowledges the fact that individuals often have multiple motivations. One important critique of this theory and participant motivations work more generally is the finding that participants often engage in online communities out of habit rather than through active, conscious decision-making \cite{wohn_habit_2012}.

Importantly, our understanding of online community participation is muddied by the imprecision of the concept of participation \cite{malinen_understanding_2015}.
When we consider motivations in online communities, it is natural to ask, ``Motivations to do what?'' For example, \citet{lampe_motivations_2010} distinguish between ``participation between users'' and ``participation directed to the whole community.'' Moreover, users participate in many different ways, such that a ``typical'' community member is hard to define: many members never post, but simply ``lurk'' and consume content, while others are active posters to varying degrees, and yet others are more deeply engaged in the work of maintaining the community as moderators or administrators. Research---especially research based on digital trace data---typically ignores the often invisible role of lurkers or those who only engage in peripheral activities like upvoting. While much of the literature on fostering participation has focused on encouraging active posters versus lurkers, some researchers have argued that lurking should not simply be viewed as passive activity but that it can represent deep engagement with a community and can encourage contributions by others \cite{antin_readers_2010, lee_lurking_2006, zhang_group_2011}.

An alternative approach in considering why people participate in online communities zooms out from the individual and focuses on the community. Work from this \emph{community manager perspective} typically measures either aspects of the technology or community-level measures of activity in order to predict community-level outcomes. For example, researchers have found that the \textit{quality} of information and content in a community substantially matters in encouraging continuous participation \cite{lu_encouraging_2011}; that group-level norms, such as those for how to deal with newcomers, can help attract and retain new members \cite{halfaker_dont_2011,lampe_follow_2005,morgan_tea_2013}; that social and technical mechanisms for managing and moderating a community can affect a community's long-term ability to thrive and overcome challenges \cite{kiene_surviving_2016, lin_better_2017,morgan_tea_2013,halfaker_dont_2011}; and that meta-characteristics of a group (such as size, activity levels, or network structures) can help predict ``success'' \cite{crowston_social_2005,crowston_core_2006,cunha_are_2019}. In this research, community growth is typically taken as a goal, and often appears as a dependent variable in regression equations measuring the effect of interventions or variations in community features on community success \cite{backstrom_group_2006, kairam_life_2012,tan_tracing_2018,whittaker_dynamics_2003,kraut_building_2012,butler_membership_2001,cunha_are_2019}. 

\subsection{The role of size in community participation}
Large communities seem to provide two main kinds of benefits through their size: (1) from higher volumes of activity, the ability to produce a sense of liveliness and maintain a flow of content that can attract and retain users; and (2) drawing from a larger pool of participants, the ability to not only obtain diverse knowledge and information from different participants but also in turn meet a broader set of needs.
Maintaining a sufficiently large and active number of participants is a fundamental concern for online communities: if a community does not have enough active members, it will not have content for others to engage with, and the community will stagnate and eventually die. 

Activity is not just a basic existential concern, but something key to stimulating further engagement in the community: contributions (e.g. posts and questions) serve as a basis to trigger further responses (e.g. discussions and answers); potential new members are attracted to a community when they observe an active community with content that matches their interests \cite{cho_how_2021,zhang_group_2011}; increased interactions between individuals via posts and comments in a community can increase an individual’s commitment \cite{arguello_talk_2006}. For example, in a study on Yahoo! Groups, \citet{backstrom_preferential_2008} found that users who later went on to become heavily-engaged in a group were far more likely to have quickly received a response to their first posting. Moreover, because participation in online communities is generally tenuous---one can easily depart a community and never come back---constantly recruiting more members is crucial. Continuously attracting new members can also create the impression that a community is lively and in turn, encourage participation from others \cite{dabbish_fresh_2012}.

As a result, quantitative metrics of online activity---such as the number of posters and commenters per day---can serve as natural indicators of a community’s success \citep{patil_predicting_2013} and are frequently utilized by researchers \cite{backstrom_group_2006, kairam_life_2012,tan_tracing_2018,whittaker_dynamics_2003,kraut_building_2012,butler_membership_2001,iriberri_life-cycle_2009,cunha_are_2019}.
A large body of research has focused on design mechanisms and strategies to stimulate growth in both size and activity. Underpinning this approach is the assumption that all online communities will try to recruit and retain members---that is, that they will try to grow. Or put another way, communities that are meeting their members' needs will grow, because newcomers will not join (and current members will want to leave) a community that is unable to meet their needs.
One of the most well-known frameworks for this kind of community growth is \citet{kraut_building_2012}’s \textit{Building Successful Online Communities}, which lays out a series of design suggestions for online community designers and managers. Derived from a rich body of research, these design mechanisms take up a top-down ``social engineering'' approach, wherein the basic assumption is that community managers can make design choices that will shape the community’s success. This and similar work often directly or indirectly proposes at least one of the following three goals: (1) increasing the number of community members by attracting new members, crafting early experiences, and socializing newcomers \cite{backstrom_group_2006,cho_how_2021,halfaker_dont_2011,halfaker_rise_2013,lampe_follow_2005,morgan_tea_2013,backstrom_preferential_2008} 
(2) retaining existing community members via strategies to increase individuals’ commitment to a community \cite{farzan_increasing_2011}; and (3) increasing contributions by and interactions amongst community members \cite{arguello_talk_2006,ling_using_2005}. 

Two important concepts that have driven this idea of growth-as-success are \textit{network effects} and \textit{critical mass}. Network effects refers to the phenomenon in which the value of a good increases with the number of users and is generally used to discuss digital platforms more broadly but translates to individual groups as well. For example, on the peer production site Wikipedia, the more users contribute to and create articles the larger and broader a repository of knowledge Wikipedia becomes and the less likely a potential contributor would choose to contribute to a rival online encyclopedia. Similarly, communities like subreddits which rely on user contributions are likely to benefit if they become the center of discussion about a certain topic.
The concept of critical mass comes from theories proposing how to attain successful collective action \cite{oliver_theory_1985} and has been applied to interactive media \cite{markus_toward_1987} and more specifically, online communities \cite{raban_empirical_2010}, as the idea that there is some key level of participation that allows a community to reach a critical mass beyond which the community will continue to grow \cite{kraut_building_2012}. 

Many of the reasons that people have to participate in online communities, from entertainment to information-seeking/sharing, seem at first blush to be more compelling in larger communities, and it seems like users would naturally gravitate toward ever-larger communities. 
Thus, in social computing literature, small communities are seen as those that are either on the path to growing or failing. Despite this, small communities exist in large numbers all over the internet. Moreover, they persist in their smallness, as shown in Figure \ref{fig:over_time}.

\subsection{Small online communities}
Given the advantages of size noted prior, why might individuals take part in these persistently small communities? One potential explanation is found by looking beyond quantitative metrics to another common framing of success of an online community: the quality and nature of the relationships fostered within it, as noted by \citet{iriberri_life-cycle_2009}. Indeed, there are some crucial motivations for participation in online communities explicitly tied to relationships that seem like they could be better met in small communities. For example, large online communities are almost by necessity less personal, making it more difficult to build `bond-based attachments' and some kinds of `identity-based attachments' \cite{ren_building_2012} such as feeling heard, feeling like part of a community, or feeling close to others.
Shared identity and dyadic social bonds can each help generate a ``sense of belonging'' \cite{backstrom_preferential_2008,ren_building_2012}, an important reason why people participate in online communities \cite{lampe_motivations_2010}. However, small and large groups each have their own advantages in fostering identity: small groups may be more likely to easily develop a cohesive shared identity than large groups while larger groups' higher activity levels could help make it more likely that an individual gets a response at all and in turn feels seen. 
Notably, small communities have a marked advantage in encouraging bonds between users: in small communities the same two individuals are more likely to interact repeatedly, fostering opportunities for deeper social and interpersonal bonds.

Alternatively, we might speculate that small communities are appealing not because they provide benefits unavailable in large communities but because they avoid the troubles that plague them. With increased membership, interpersonal conflict becomes more likely and more difficult to resolve \cite{tausczik_impact_2019}, as do undesirable contributions such as spam and posts or comments by trolls or newcomers who don't know the community rules \cite{kiene_surviving_2016}.
At times, the governance work of these communities requires a substantial amount of time, coordination, and labor across a team of moderators \cite{blackwell_when_2018, roberts_behind_2014}. 
Larger communities typically deal with greater heterogeneity amongst members and must deal with attempts to shift the boundaries of their community---both topical and normative---in ways that generate intra-group conflict \cite{butler_cross-purposes_2011}. All of these issues fundamentally influence the types of experiences, interactions, and relationships participants might have in a community, and small communities have a distinct advantage with respect to community management.

However, it remains unclear what benefits are generated \emph{uniquely} by small communities. Research has rarely explicitly sought to understand how online communities or participant motivations differ by community size. A few recent studies suggest that small communities may differ in important ways. For example, \citet{tan_all_2015} found that participants on Reddit ``post increasingly evenly among a more diverse set of smaller communities'' over time. In a study looking at wiki communities,  \citet{foote_starting_2017} found that many community founders in fact intentionally start small communities, focused on narrow topics, without seeking or expecting a large following.

While the growth-as-success assumption underlying much of social computing research on online community success and participation suggests---and at times explicitly states---that small communities are failures, the fact that they not only exist in great numbers but also persist over extended periods of time suggests otherwise. Previous research sometimes uses community size as a variable to predict future activity levels, but we know precious little about the ways that participants perceive and experience communities of different sizes, or how size plays a role in participants' decisions to participate at all.  Better understanding when, how, and why participants choose to take part in small communities that persist can help to moving towards a more nuanced understanding of community success and the ways in which participants diversely find value in different kinds of online communities.

\section{Approach and methodology}

\begin{figure}
    \centering
    \includegraphics[width=.8\textwidth]{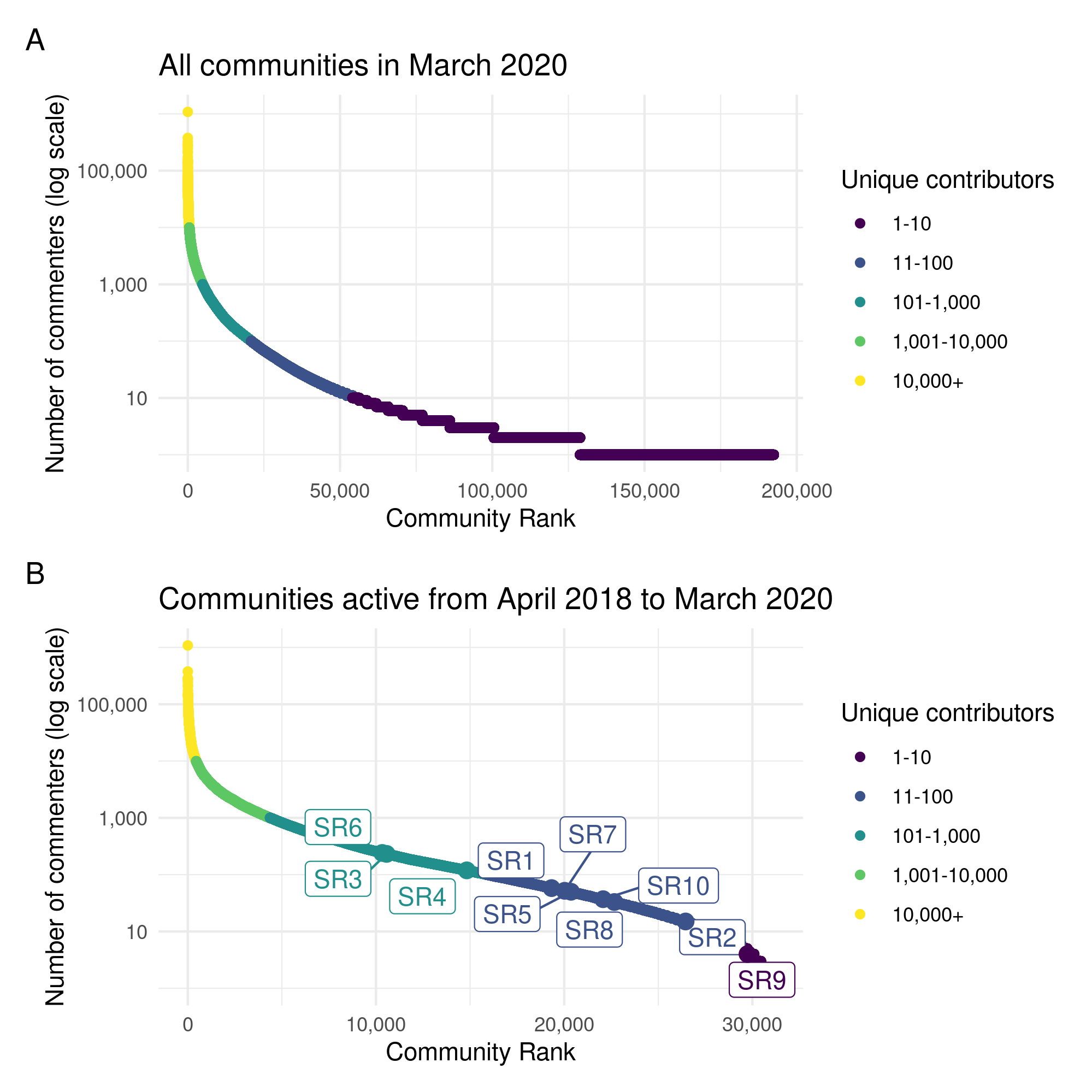}
    \caption{These plots show the number of unique commenters who commented on each subreddit in March 2020. \textbf{A} shows all subreddits that had at least one comment. \textbf{B} shows only those subreddits which had at least one comment in each of the previous 23 months. This filtered list included our sample for interviews, and subreddits that we drew participants from are are highlighted.}
    \label{fig:long_tail}
\end{figure}

The site of our research is Reddit, a popular discussion-based website made up of over one million communities called subreddits.
Reddit's subreddits can be seen as the contemporary equivalent of classic online communities like those on Usenet, and each has their own topical focus, moderation team, and subscribed users. Subreddits have been the focus of a large amount of research from online community scholars, examining both individual subreddits \citep{kiene_surviving_2016,dosono_moderation_2019, laviolette_using_2019} and the relationship between them as they sit in what \citet{jhaver_did_2019} call a ``multi-community environment'' \citep{jhaver_did_2019, chandrasekharan_internets_2018,fiesler_reddit_2018}.

Reddit is particularly well-suited for studying communities of different sizes, with subreddit communities that see monthly contributors in the single digits to those in the hundreds of thousands---capturing a wide range of ``smallness''. Its advantages as an empirical site of social computing research additionally include its popularity and cultural influence as well as the publicly accessible data it provides. Because of this, Reddit has been extensively examined in social computing and CSCW literature \cite[See][]{baumgartner_pushshift_2020}. A number of large-scale studies explore things like the dynamics of participation \cite{gilbert_widespread_2013,tan_all_2015,tan_all_2015} and the creation and spread of moderation practices and rules in communities \cite{chandrasekharan_internets_2018,fiesler_reddit_2018}. Qualitative research has explored the work of moderation \cite{dosono_moderation_2019}, gatekeeping \cite{leavitt_role_2017}, and the dynamics of newcomer integration \cite{kiene_surviving_2016}. Design work has explored topics like how interface changes can influence cooperative behavior \cite{matias_preventing_2019}.

\subsection{Defining a set of small communities}

To answer the question of why persistently small online communities exist, we must identify what constitutes a small community on \textit{Reddit}. 
Notions of `small' and `niche' are contextual, relative, and likely differ between platforms, between technologies, and between users. For example, a group text starts to feel large once there are ten or twenty active participants, while a `small' subreddit may contain dozens or even hundreds of participants and even more lurkers.

In Reddit, as in many online contexts, we see extremely skewed patterns of participation and attention. The top panel in Figure \ref{fig:long_tail} shows the number of unique users who commented in each community on Reddit in March 2020. The vast majority of subreddits have very few participants, while a few have tens or even hundreds of thousands of unique participants each month. In our study, we locate `small' communities in the long tail of participant counts seen in Figure \ref{fig:long_tail}.

Two types of communities are temporarily `small': those that are on their way to growing larger and those that are on their way to disappearing. While more research is needed to understand why users participate in extremely small and extremely short-lived communities, this project focuses on understanding what we are calling \emph{persistently small communities}. As Figure \ref{fig:over_time} shows, most consistently active subreddits remain approximately the same size over time, and most of these subreddits typically have between 10 and 1,000 monthly participants.
For this study, we used the Pushshift Reddit dataset \citep{baumgartner_pushshift_2020} to identify a subset of persistently small subreddits, which had between 5 and 100 unique contributors monthly for at least 36 different months and under 10,000 subscribed members.\footnote{This is slightly different from the filtering criteria used to produce Figures \ref{fig:over_time} and \ref{fig:long_tail} but produces a very similar set of subreddits}

\subsection{Data collection}
We conducted 20 semi-structured interviews with participants of Reddit to learn about their experiences on small subreddits. The interviews were conducted in two rounds, one in November 2020 (twelve interviews) and one in February 2021 (eight interviews). The interviews were transcribed within a week of the interview but analyzed after each respective round of interviews. At each round, this allowed us to look across individuals' experiences, develop inductive codes, and reflect on how the interviews related to prior work on participation and community size.

Participants were recruited through posts on the subset of small subreddits identified, as described in the prior section. We used a simple script to generate a random set of 100 communities to begin. We then performed a manual inspection of the sample of communities, with the goal of ensuring an final sample that fulfilled two criteria: (1) online communities across a variety of (small) sizes; (2) online communities focused on both hobbyist interests and non-hobbyist interests. To ensure that we covered these, we followed the non-representative stratified sampling method described in \cite{trost_statistically_1986}.\footnote{A table outlining our approach is included in the Appendix.}

As part of our inspection, we also ensured that the community was active (and not, for example, simply populated by automated posts). Once a community passed each of these checks, we reached out to moderators via the modlist\footnote{This is a messaging system that sends a message to all of the moderators for a given subreddit.} to ask for permission to post in their community. Upon approval, we posted a recruitment post with a link to a screener survey and then scheduled interviews. Following interviews, we created transcriptions of each interview. 

In total, we reached out to 42 subreddits. In the end, we interviewed 20 people recruited from 10 unique subreddits, two of which were subreddits aggregating research opportunities that we did not directly post to, but where others shared our recruitment post. Because those subreddits met our size and activity criteria, we kept these interviews in our dataset. Furthermore, in the course of our interviews, we also discussed other subreddits that interview participants were active in. Metadata about the primary subreddits and the subset of these secondary subreddits that are explicitly cited in our findings appears in Table \ref{tab:sr_description}.

Each interview ran between 45 and 90 minutes and was conducted in English. Participants were given \$15 USD for their time. All interviews were conducted by the same interviewer (the first author) via password-secured video calling. The interview protocol we used to guide these interviews focused on: (1) the participant's experiences in the small community they found our post on; (2) other online communities, Reddit or otherwise, the participant was active in; and (3) comparisons between the experiences in the small community and other communities. We sought to contextualize each participant's experience of the relevant small community within their broader online experiences in order to better understand the value the small community provided.

\subsection{Description of interview pool}
\label{section:interview_pool}
An important note about Reddit is that its user base is heavily skewed, with users usually being younger, male, and based in the West and Global North \cite{sattelberg_demographics_2021,pew_research_center_who_2019}. While all of our participants did come from the West and Global North, our resulting interview pool was much less skewed in other respects. The interview pool consisted of 20 individuals. We did not interview minors, and participants ages ranged from 18 to 48 with a median age of 30. Participants were located in three different countries, with 18 being in the United States and two in Europe. Within the United States, participants were from 12 different states. Our participants included 10 men, eight women, and two non-binary individuals.

Participants were recruited from 10 unique subreddits. Their tenure on the subreddits ranged widely: one participant had joined their community earlier in the week and others had joined within the last year, while some participants had been in their community for over 10 years.

The subreddits that participants were recruited from vary in their degree of smallness: in March 2020 (the last month for which we have full data), they saw anywhere from 5 to 240 unique contributors---all small relative to Reddit, but variable in \textit{how} small. We provide a description of the original ten subreddits and three other small subreddits substantively discussed in the interviews in Table \ref{tab:sr_description}.

\subsection{Qualitative analysis}
Our analytical approach followed Braun \& Clarke's thematic analysis approach \cite{braun_using_2006}. The analysis process involved two stages, corresponding to our two rounds of interviewing. After the first round of (twelve) interviews, the first author performed an initial line-by-line inductive coding of all the interviews, iterating over the interviews as needed. The first author then organized codes into a visual ``map'' of related codes, iterating over them to cluster and organize codes based on the transcription content they coded. The first author produced an initial set of memos to go with this code map, and then both authors reviewed it to reach agreement on the codes. 

Guided by the map, we generated memos across all interviews to identify themes and patterns in the interviews as well as connections across codes that might respond to our guiding question of the value participants found in the small communities they participated in. Memos were generated in multiple rounds, from continued discussions between the authors and after receiving feedback from colleagues in our research working group. Guided by our initial questions about value and small communities, our attention leaned to how and why participants distinguished the community from others and in what ways they were engaged with it.

From this first set of interviews, we found that we did not achieve information saturation \cite{corbin_basics_2015}. Based on the gaps and further directions identified in the memos, we adjusted our interview protocol and conducted our second round of interviews. After the second round of interviews, we again performed line-by-line coding for the eight new interviews. In this round, we primarily utilized the codes from the first round of interviewing and coding but also generated new inductive codes as appropriate. We updated the code map and wrote more memos, which focused on synthesizing across both sets of interviews. Leveraging two rounds of feedback---one on memos and one on initial findings---with our research group, we held several rounds of discussion together about the memos and code map to refine the findings until reaching agreement.

\subsection{Anonymity and privacy of participants}
Because Reddit consists mostly of public pages as well as viewable histories of a user's activity, it could be possible to identify a user, especially those in smaller and topically-unique communities. As a result, we modify, obfuscate, or omit details that could either identify the participants, specific users, or the subreddits in question. The stories and quotes below often reflect points and issues raised by multiple interview participants.

\begin{center}
\begin{table}
\begin{tabular}{|l|l|l|l|l|l|}
\hline
\textbf{SR} & \textbf{Pseudonym} & \textbf{U.C.} & \textbf{Age (year)} & \textbf{Topic} & \textbf{Participants} \\\hline
SR1 & data & 60 & 6 & data archiving & P1, P3, P7\\ \hline
SR2 & manga & 15 & 10 & a discontinued manga & P2\\ \hline
SR3 & visions & 230 & 12 & visions of the future & P4, P5\\ \hline
SR4 & fiber & 120 & 12 & a fiber art technique & P6, P9, P12\\ \hline
SR5 & band & 50 & 9 & an inactive music band & P8, P11\\ \hline
SR6 & breakfast & 240 & 11 & a breakfast food item & P10\\ \hline
SR7 & friends & 50 & 6 & a group of friends& P13, P17-19\\ \hline
SR8 & research 1 & 35 & 4 & paid research opportunities & P14, P15\\ \hline
SR9 & beverage & 5 & 7 & exchange of beverage ingredient & P20\\ \hline
SR10 & research 2 & 35 & 4 & paid research opportunities & P16 \\ \hline
SR11 & influencer & 740 & 5 & mocking a social media personality & P2\\ \hline
SR12 & pregnancy & N/A\textsuperscript{\textdagger} & N/A\textsuperscript{\textdagger} & pregnancy discussion & P19\\ \hline
SR13 & pet & 20 & 8 & an unusual pet animal & P8\\ \hline
\end{tabular}
\caption{Descriptions of subreddits (SR). SR1-10 are the subreddits recruited from. SR11-13 are additional subreddits mentioned that come up below. U.C. stands for ``unique contributors'' as of March 2020 (rounded to the nearest 5).\\
\textsuperscript{\textdagger} Subreddit is private and data is not available.
}
\label{tab:sr_description}
\end{table}
\end{center}

\section {Findings}

We identified seven major themes from the interviews we conducted. The first four directly consider the benefits that our participants receive from their small communities. A fifth theme considers how participants conceived of and perceived the size of the community, while the final two themes consider how small communities relate to other communities that participants participate in.

\subsection{"Singularity of purpose"}
Small communities often provided little pockets of highly specific content. This hyperspecificity of topical focus of the community is a major factor that permeates every aspect of the community. 
Frequently, participants seemed to join and follow communities that appealed to a specific purpose or goal they had in mind; that is, they specified narrow slices of content types on a topic by following smaller communities that were offshoots of larger ones covering the topic more generally. For example, SR1 (data) deals with data archiving. This community is in essence a sub-community to a larger community of data-interested individuals, but SR1 focuses on a specific task that must be done in data archiving. Similarly, within the fiber arts community, SR4 (fiber) focuses on a particular method for a fiber art technique and is an offshoot of a large community centered on that technique more generally.

Even if the title of the community does not sound narrow, the way that the community approaches the topic is specific. In order to preserve anonymity, we give a hypothetical example: a subreddit called `r/makeup' might focus on reconciling with significant others with no posts about cosmetics. SR3 (visions) has a similarly ambiguous title, but posts focus exclusively on only one meaning. Similarly, SR8 (research 1) and SR11 (influencer) are both communities whose topics could be construed in multiple ways. Yet, each focuses on a specific aspect of their respective topic: SR8, about research participation opportunities, focuses on only specific \textit{types} of opportunities; SR11, about a social media personality, chooses to use the space specifically to cyberbully him. The hyperspecificity of all these communities point to a ``singularity in purpose'', described by P19 as generally important for communities to have to succeed. It is through narrowness that participants were able to glean what belongs in the community and develop clearer expectations about the content.

Moreover, this narrowing of the space leads participants to believe that other members had a high level of emotional investment. As described in later sections, several participants described how the community is so niche that people who are not that invested in the singular purpose of the community do not show up. Of course, not all participants were so engaged with the communities they discuss and not all communities were so involved. For example, P4 and P10 described their communities (SR3 (visions) and SR6 (breakfast), respectively) as something ``just interesting to me'', and ``fun'' while simultaneously noting that their engagement with the community was ``lighthearted'' or ``just for entertainment''. However, even in these cases, the communities maintain their hyperspecific nature.

\subsection{Seeking expertise from narrow niches}
Participants described how joining these small, hyperspecific communities helped them understand \emph{who} they are interacting with. For example, P15 stated that people on Reddit were there for ``the serious version of what they're looking for'', and that on SR8 (research 1) he knew that he could take people seriously and trust the content there.

Similarly, P3 noted that SR1 (data) is made up of a ``certain type'' of person interested in making progress towards solving a problem that's central to the community's topical focus. 
In other communities, the topic is niche enough that the only individuals who go there are those who care enough about the topic to be discussing it in a serious manner.
Thus, participants expected information and content in these communities to be of high quality and to find space for ``serious'' discussions. Moreover, they frequently expected expertise on the topic from the community, particularly in hobbyist subreddits, that they do not expect from larger communities:

\begin{quote}
    [...] 
    I can probably make a safe assumption that people there more often than not know what they're talking about. I'll definitely be much more specific and not try to water questions down with like, my broader scheme of things---I can get as technical as possible, right? If I were to ask like the same question over at [the larger parent community], I might want to give a little bit background on what I'm trying to do, why I'm trying to do it, you know, other things that I'm using, but [in SR1], I can just be like, hey, look, I'm trying to use this algorithm for this one thing. Why should I? Or should I not do it for this? (P7)
\end{quote}

The expected expertise makes the community not only relevant but also more efficient for information-seeking and advice. These attributes both contribute to a distinct sense that the information in the community is reliable, or as P7 expressed above, he ``can probably make a safe assumption that people there more often than not know what they're talking about.'' This holds true even for less explicitly hobbyist subreddits like SR6 (breakfast); P10 said of SR6: ``[...] if I ever have a [food-item] related question, they'll jump right on it.” Moreover, the community is sometimes the only space that a participant can get expertise on a specific topic. In discussing SR13 (pet), P8 remarked that there is nowhere else she could get similarly detailed information about how to take care of her uncommon pets.

On the other hand, participants noted that it was relatively easy to spot an `outsider' who didn't actually care about the topic the way that community members did. P11 discussed how he and the rest of the community in SR5 (band) valued authentic interest in the the band that was the focus of the community. Specifically, he could tell whether someone in SR5 actually cared about the band: 

\begin{quote}
    The people who are fanatical about it, and like know every word to every song [by the band] and stuff... It's obvious who those people are, and it's obvious what they're getting out of it. And you can tell when somebody is not functioning on that [level]. I'd say that there's more often than not, there's a---that personal revelation kind of subtext going on in a post, somebody posting about the genuine value that this album has brought to their life.
\end{quote}

Put another way, because individuals in the community are interested in the highly-specific topic of the community, this allows participants to bypass noise and engage in rich discussion about it. In this way, homogeneity of interest-level actually enables diversity along other dimensions and participants frequently described how \textit{diverse} the perspectives, content, and people in their communities are. For example, one of the things that P3 appreciated about SR1 (data) was the opportunity to hear other peoples' perspectives and contribute one's own perspectives on how to approach a data problem. Pulling up an example of a post he enjoyed in SR1, a discussion of a technological tool, P3 noted: 

\begin{quote}
    [...] you can see that there's people talk about the pros and cons of [the technical tool]. And for me, the discussion is kind of fundamentally [affirming others' points] and then giving a different viewpoint, right? Where it's like, \textit{oh, it's slow. It's difficult to use}. It's like \textit{yeah, I echo that}. And then you have the personal feedback of \textit{Hey, I haven't seen this mentioned, but I had an issue with [the tool] rotting or scratching. And oh, by the way, here's a cool different way to deal with it, if you want to take a look at that}. So it's kind of like trying to share a mixture of experience along with personal pros and cons.
\end{quote}

\subsection{I care about the personal, but not the person}
One way a community can provide distinctly unique, original content is through posting personal experiences, information, and content contributed by users. For example, P5 follows SR3, a subreddit related to discussing predicted visions of the future; she is unlikely to find a group of people engaging in this and sharing their personal visions in other places. Several participants in hobby-related subreddits mention the frequency of posts showing off someone’s craft. One particularly interesting aspect to the personalized nature of the unique content is also the \textit{proxy experience} that consumers of that content get. P8 described how she is able to “experience” the band she likes through the stories and recordings posted by community members, especially valuable because the band is permanently disbanded following the death of the lead singer. P3 and P7 talked about learning from the projects of others or benefiting from the labor-intensive projects done and shared by people on SR1 (data) that they find interesting but would not pursue themselves.

Given the importance of the \textit{personal} in the content of these communities, one might assume that participants in these small communities come to know one another and build relationships. However participants repeatedly noted that they do not particularly have an interest in individual users nor do they notice them. Participants often shrugged off questions about other users with comments like, ``I just don't really tend to pay attention to [usernames]'' (P11) or ``I don’t really look at the names'' (P16); participants were typically more interested in interacting with the group as a whole rather than in one-on-one interactions with individuals. Participants reported maintaining a distance from other group members and engaging primarily with the content and not the poster. At the same time, participants strongly believed that others in the community are ``like them'' because they share this very specific interest, suggesting a sense of shared group identity amongst community members. A few participants discussed how finding people who were like them made them feel like their interest was more “normal”. Sometimes this was through finding others who had an even more extreme interest in the topic, like the case for P10 on SR6 (breakfast):
\begin{quote}
    I like [food item] a lot more than most people [...] a lot of times what happens when you get introduced to a subreddit that's related to an interest you think you're at the extreme end of interest, and then you realize that there are people who go so much deeper than you. So it makes you feel a little more normal as well. Because there are people who, you know, they post like their [food item] hauls, and you know, they'll have pictures of dozens of boxes of [food item], like obscure things I've never heard of [...]
\end{quote}

Although uncommon, we did find cases where participants personally know other users, beyond Reddit, and even made some ``Reddit friends''. P1 talked about his friend who introduced him to the data archiving community. P9 and P12, both in two fiber arts communities (one being a large community), noted that they know some of the people they interact with on Reddit, though it’s not clear how they got to know these people. In the case of fiber arts communities, there is a strong online community across multiple platforms as well as offline events.\footnote{Ravelry is a major platform used by members of fiber arts communities and frequently mentioned by participants.} Despite these exceptions, nearly all of our participants---including P1, P9, and P12---described focusing on the content of the community rather than on the individuals who participate in their respective communities.
The only true exception was SR7 (friends). This community is composed of a group of friends, many of whom met each other through the subreddit for the city that they live in. Our other interview participants suggested that this is rare and that even most small subreddits are composed primarily of strangers who remain strangers.

While friendship formation was rare, a few participants noted that there are some usernames they come to recognize over time. These users are sometimes users who are very active across many subreddits, though participants did at times recall specific users who had made some memorable contributions to the community (although even in these cases they tended to have difficulty recalling the specific username). For example, P7 in SR1 (data) recalled a user who had made a series of highly detailed posts about a massive data project because they enjoyed the posts. However, P7 went on to explain that he isn't particularly interested in the individual user, just the content.
In fact, several participants noted that to be recognized on Reddit often carries negative connotations: a widely known user is probably infamous rather than famous.

\subsection{Getting away from negativity}
Overall, participants saw their small communities as positive, supportive spaces. They often described their communities as ``different'' from the rest of the Reddit or other spaces on the internet, which they described as ``toxic''. For example, P11 called SR5 (band) ``one of the few places on Reddit that I genuinely feel like I found something that feels different [from Reddit and the internet more broadly]''; P5 noted that she enjoys being in SR3 (visions) because she can just ``get away from all the negativity [in the world].'' 

Repeatedly, participants described the goals and motivations of others as being more pure in small communities. This was often described in contrast with larger communities where others often post or comment ``for Reddit karma'', which is a sort of points system that tracks popularity and engagement for each user. Because there are so few users in small communities, participants claimed that those just seeking karma go elsewhere. In fact, participants frequently noted that they did not post unless they felt they had something useful to contribute. Many participants were self-declared lurkers who preferred to consume the content of the community. However, participants did not indicate that they felt that being a lurker meant they were not part of the community. For example, many participants actively up- and downvoted posts---which they saw as an important way to support other individuals' posts---and several frequently mentioned at least skimming through every single post.

Participants described their communities as spaces where dialogue, discussion, and conversations can be held in productive, positive ways. Participants frequently referred to conversations as respectful or civil; described other members of the subreddit a ``good people'' (P13) and generally ``nice'' (P5); and noted that actual dialogue and discussion could be exchanged in the community without devolving into harassment and arguments (in contrast to other online spaces).
For some subreddits, particularly in SR1 (data), where there is a distinct task being addressed, multiple participants described these interactions as a kind of collaboration amongst community members. 
More generally, participants described a sense of support felt in the community and a desire to hold it up, as the community served as a unique space for this topic. 
This sense that the community is a unique space of discussion---not only topically but also as a positive, uplifting space---is important because it additionally can make some content that might be regarded as repetitive, annoying, or boring now acceptable. For example, P6 noted of his fiber arts community: 

\begin{quote}
    Well, the not so interesting part is usually when stuff is like, \textit{I've already tried it before}. But at the same time, you can keep it interesting by reminding yourself that you're someone who's got a little more experience maybe than the person who's posting, you can offer your advice, how you've conquered specific challenges before, you know, that sort of thing. So [...] just because a post in and of itself isn't intrinsically interesting, doesn't mean you can't get something interesting out of it.
\end{quote}

In turn, negative incidents are seen as being committed by outsiders. P11 described a time where a poster seemed to try to essentially scam the members of the community. When asked if the incident had changed how he felt about the community at all, P11 dismissed the notion, stating that it was clear that the poster could not have been a ``real'' member of the community. While talking about the lack of conflict and problems within his community, P7 described this as everyone having shared expectations: ``When you have a really small group of people [...] it's more likely for everyone to be on the same wavelength.'' Moreover, the narrowness of the topic that the small community centered around also helped ensure that participants were ``on the same wavelength''.

Part of this impression may be due to the fact that the small communities manage to self-moderate without much apparent governing overhead. Participants often mentioned never noticing moderators or moderator action, describing them as rather invisible and moderation as ``bottom-up''. Moreover, participants were frequently unaware of what the rules of the subreddit even were, or reported that the community actually had no subreddit-specific rules. 

Notably, P20 provided a counterexample to the effective self-moderation narrative, recounting difficulties in handling spam bots in a small subreddit he started and moderated, an issue which P20 believed arose after bots started targeting smaller communities attached to larger ones (as the subreddit in question was an offshoot of a broader community). The spam bots became such a problem that he had been forced to ``purge'' the posts in the subreddit, after which he found that ``nobody had really posted anything there [since].'' However, P20 also noted that this was an exception and bots are usually less of a problem in small communities in general: ``I think if you're in a smaller community there are fewer bots [...] and there's more actual interesting posts being made as a percentage [of all posts in the community].''

While our participants overwhelmingly described their experiences in smaller communities as positive and supportive, obviously not all small communities are wonderful, happy spaces. Indeed, it is possible that finding a community of like-minded others can at times encourage hateful or dangerous behavior. For example, while talking about SR2 (manga), P2 briefly described SR11, another small community in which he had participated in the past and which was dedicated to mocking a social media figure with a disability. However, even though those norms promoted horrific behavior, it is worth noting that individuals still held shared expectations about how to act, what to post, and what might be considered acceptable. As a result, even in this toxic community there was a lack of intra-community conflict, similar to the other more pro-social communities in our dataset.

\subsection{Awareness of size and ambivalence about growth}
How did participants in small communities think about the size of the community and community growth? Size in general appeared salient to participants, who typically conceived of the communities that they belonged to as small, and often described experiencing qualitative differences in communities of different sizes, some of which are captured above.

Some participants explicitly stated that they liked the smallness of their community: this way, the rate of content was reasonable such that they could read or skim all of the posts and uninteresting spam didn't make its way into their feeds. Moreover, the small community size contributed to a sense of shared understanding about what the community was about and helped to protect the community from trolls, spammers, and karma-seekers, who would get little pay-off from harassing a small community.

However, size was not necessarily at the forefront of participants' minds. Participants demonstrated a mixed bag of feelings about size as a characteristic of the communities they engaged in. There were those who didn't care about size at all (one participant remarked, ``I literally don't think about it at all when I look at subreddits''), those who recognized it as a factor in their community experience, and those were very aware of the sizes of the communities they engaged in. Frequently, the third group was made up of individuals who were more involved in Reddit, such as long-time Reddit users or moderators on the focal community or elsewhere.

Across these groups, participants varied in the degree to which they desired growth. Most participants expressed neutral sentiments about growth: they neither minded it nor saw it as a negative thing, so long as the existing character of the community would be maintainted. In part, these participants did not mind growth because they believed that even with growth, the community was topically unique enough that they did not anticipate many problems: either the community would hit its natural size limit due to the scope of the community's topical focus or the topic was specific enough that the only people joining would be those who care about the topic.
A few participants actively wanted to see growth because they wanted to see more content from the community. P6, for example, noted that he actively commented or upvoted posts and comments because he wanted to see the community (SR4, fiber) survive. He encouraged individuals to post and pushed for the community to grow enough to reach ``critical mass''. However, participants like P6 also frequently noted that this growth was undesirable in sudden bursts; P1 said he would like for SR1 (data) to grow, but only in a `trickle'.

Overall, size was salient, but smallness did not present a major concern for participants and at times was seen as beneficial. There was no clear call for growth across participants, suggesting that growth is not a central desire for participants in our set of persistently small communities.

\subsection{Curating an online experience}
\label{section:curating_online}

Conversations with participants revealed that following small communities was not only a way to be exposed to content about a topic, but also part of a larger strategy for curating a personalized experience across all of Reddit.
All of the interview participants took part in multiple communities, some being subscribed to hundreds of subreddits. Thus, their participation within these small, highly-niche communities is part of a broader strategy of curation of content across communities. One participant, for example, on a subreddit about research study opportunities, noted that all of the subreddits she follows are school- and career-related. As a college student, she uses Reddit to get the latest updates about her school as well as potential career paths she is considering taking. In this vein, her participation in the research studies subreddit is in part related to her interest in conducting research in the future.

Other participants tended to have a broader range of subreddits they followed (i.e., not necessarily just work-related). However, the basic principle of curating for specific content of interest remains consistent. While participants are intentional about why they continue to subscribe to and participate in subreddits, they often described initially joining communities without much active thought. They described finding subreddits in various ways, often saying that they joined the subreddit almost immediately or didn't have to really think about it before clicking ``subscribe'' as it aligned with their existing interests. 
Joining and leaving communities is part of a continuous process, where subreddits are continually added and removed from a stream of content.

At times, participants engaged with both the small community as well as a larger parent community that engulfed it. Small communities discussed by participants were frequently \textit{nested niches}: offshoots of existing, larger niche communities. Even in cases where they were not direct offshoots, participants still typically discussed their small community in relation to larger, broader communities on related topics.
The preference in following these smaller communities nested within larger, more general ones results from a number of motivations. First, some participants noted a higher signal-to-noise ratio. Smaller communities avoided repetitive content that could get frustrating, not only because fewer people posted but also because the smaller community filtered out content uninteresting to the participant, because it was focused only on the subspace of that topic. For example, P19 noted how SR12 (pregnancy) was an offshoot of a larger community. Both are about pregnancy, by SR12 is specifically for expectant mothers delivering in a given month. In fact, the larger community links to several communities like SR12 for different due dates. When asked why SR12 was preferable to the community it is nested within, P19 stated:

\begin{quote}
    [...] it's more condensed. You know, because it's, I feel like it's on a shorter, tighter, smaller timeline. [In the parent community] you get in different cohorts of people coming in and out, so they ask the same questions. I didn't realize that these questions were asked a billion times over and over in a different way, by different people. But now I see that happening because I've been there long enough. [SR12] doesn't annoy me as much.
\end{quote}

Other participants described wanting to avoid spamming individuals in larger communities with their specific questions, noting that they would get a better response in the smaller community anyway. Yet other participants found that the smaller community provided a more welcoming space to discuss the specific subtopic that they wanted to talk about.

Importantly, not all of the communities were explicit offshoots or nested in larger communities. However, even if a community was not an obvious offshoot of a larger one, it was defined by a distinct particularity within a broader topic. The uniqueness of content in a given community is a consistent thread throughout participants' experiences of their small subreddits. For example, P8, who is on SR5 (band), noted:
\begin{quote}
    ``[...] probably there are subreddits for just indie groups or something like that as a whole. But if you were to go into one of those, you probably get lost a bit in all of it. Like you wouldn't get [...] as well of a catered responses that you'd get in [SR5]. Like, if you went into an indie one and went, does anyone know this album, like by this? And where you can get it from? Like, they'd be like, oh, I've never heard of that band.''
\end{quote}

\subsection{Strategies for interacting with small communities}
\label{section:strategies}
Participants described a number of strategies for interacting with small communities. The main Reddit homepage surfaces content from all of the subreddits that user subscribes to, and participants described occasionally running into a new post from the small community when scrolling through their home page. However, they frequently described deliberately seeking out content from these small communities. Because small communities see lower rates of content and levels of engagement compared to large ones, these communities were---as multiple participants put it---disadvantaged in the Reddit main feed algorithm, which participants claimed favors posts and content from larger communities that have more comments and upvotes.

One obvious and straightforward way of interacting with small communities is going directly to the community page, which many participants did. They found that directly navigating to the subreddit, sorting posts by newest, and then seeing what posts they had not yet read but looked interesting was an easy way to see what they had missed. This is very different from how participants described typically consuming content from larger communities, and this approach is only possible because small communities have a slower rate of new content, so users are able to parse through all the posts and find where they last left off.

However, some participants described a more elaborate system of consuming curated content. One involved using the multireddit feature on Reddit, which allows a user to create a group of subreddits to see content from all of the subreddits included. P20 most explicitly described using this feature, explaining that the small community he was on (SR9, beverage) was part of a multireddit he created containing subreddits he was interested in related to homemaking (cooking, food, storage, etc.).

Another strategy involved the use of more than one Reddit account. The multi-account dynamic appeared when participants wanted to take steps to protect their real-life identity or separate different parts of their lives. For example, P19, who is on a small subreddit where she interacts with people she now knows in real life, discussed the separation of the account used for that community and other personal interests and an account specifically for moderation work on addiction recovery subreddits. While she admitted that people who knew her could likely put the two together, she preferred to maintain separate accounts for these two aspects of her life. Another participant (P1) mentioned interest in having separate accounts for less identity-concerned reasons: being subscribed to so many subreddits, he mentioned that it would be nice to have a different account for subreddits he ``actually'' cared about, such as SR1 (data).

\section{Discussion}

From discussions with participants on their experiences in small subreddits, our findings support, extend, and call into question various aspects of prior work that considers community size and motivations for participation.

\subsection{Partitioning the information landscape}
We found that participants frequently engaged in information-seeking behaviors in their small communities, looking for others' personal experiences and knowledge on a topic. While information seeking is a common and important motivator for online participation \cite{lampe_motivations_2010,lu_encouraging_2011}, this finding was surprising as large communities are expected to have advantages in meeting information needs: larger communities not only have more potential sources of information but also a more diverse set of participants, both of which are important in attracting and retaining community members \cite{arguello_talk_2006,backstrom_preferential_2008}. However, participants found that small communities often served their needs better: participants noted that seeking expertise, perspectives, knowledge, and stories was more efficient in the small communities they were discussing. The narrow goals of the community meant that community members tended to be passionate and knowledgeable about the highly-specific topic of the community. In turn, participants---active posters and lurkers alike---were able to develop clearer expectations of what kinds of content and interactions belonged in each community. As a community grows, the growing flow of posts and comments to go through leads to a sense of information overload \cite{jones_information_2004}, which may outweigh the benefits of having more people and actually make some kinds of information-seeking more difficult. Our interviews indicate that by partitioning via smaller communities and curating which of these narrower communities to follow, users can manage this flow of information and expectations. 

In addition to obtaining information, users might derive certain personal and social benefits from \textit{contributing} information and knowledge \cite{jin_why_2015, lampe_motivations_2010, zhang_group_2011}. To this end, smaller communities may similarly help contributors partition what spaces are appropriate for sharing certain information. For example, P20 noted that SR9, about exchanging an ingredient to make a particular beverage, was useful because the broader subreddit about the beverage in general did not care about making the beverage or exchanging ingredients, instead finding posts along those lines potentially irritating. In other words, just as users can manage the flow of information they encounter by curating smaller communities, they can also manage what audiences they provide information for. 

This ability to choose a community that mirrors the content you want to share may be important to motivating users to contribute. While \citet{zhang_group_2011} found that a larger audience motivated greater contributions from Wikipedians, our findings suggest a deeper story. Depending on the kind of content shared, or the type of experience sought, users may actually prefer a smaller, more predictable audience.

\subsection{Control of online experiences}

The ability to select very specific topics, decide who to interact with, and manage the flow of content via these small communities indicates that an important aspect of curating content consumption for users appears at least in part to be a matter of \textit{control}. Given the scale of the internet---and a widespread sense of malaise with online hate, toxicity, and harassment---it is possible that controlling content and interactions is more important to users than ever: several participants described their online communities as special spaces to get away from the negativity on the rest of the internet. Instead, these communities provided content that participants knew they would appreciate and enjoy and which distinguished them from the rest of the online world.

Small communities may provide this controlled space in two ways. First, small communities may more easily provide positive initial interactions for newcomers, which have been found to be important predictors impressions of and long-term participation in a community \cite{cho_how_2021,halfaker_dont_2011,backstrom_preferential_2008}. P6 noted that in his fiber arts community, he felt inclined to be more encouraging, responsive, and understanding of newcomer posts even if they were repetitive or slightly off-topic because the community was small and didn't see too many posts a day. While larger communities might have an advantage in more potential respondents to engage with newcomer posts, small communities may be more lenient to newcomers not only because the volume of newcomers is more manageable but also because the narrower topical focus of small communities automatically tends to filter out users and posts that might be poorly received. Second, in general, smaller communities are easier to manage and see less undesirable contributions such as spam and trolls \cite{tausczik_impact_2019,kiene_surviving_2016}. Our interviews reflected this, as several interview participants described their communities as effectively self-moderating without the need for administrative oversight, both remaining generally civil and staying on topic. More broadly, for users, the high-specificity in content matter of small communities additionally provide users a more granular level of control in what content they choose to expose themselves to.

\subsection{Dyadic bonds and group identity}
While prior literature suggests that relationship building might be a key benefit of small communities, and one explanation for their persistence, we did not find this. Lampe et al. \cite{lampe_motivations_2010} distinguish between participation between users and participation with the group as a whole---we saw much more evidence of group-based participation and only rare instances of dyadic interactions or relationships. Participants frequently mentioned that they were not interested in actually knowing the other individuals in the group, instead engaging with the group as a whole. Somewhat confusingly, many subreddits often included personal and intimate posts, which our participants enjoyed. Even this content seemed to support and encourage group identity and a sense of belonging rather than a desire for getting to know people as individuals. In other words, although small communities might be intuitively expected to facilitate interpersonal bonds---particularly dyadic ones---participants maintained a distance from others and did not typically interact with fellow group members one-to-one.

This tension between wanting the personal but maintaining a distance from individuals might be explained in part by the fact that Reddit is a pseudo-anonymous platform, which affects they ways people approach their use of it \cite{ammari_understanding_2015,leavitt_this_2015,de_choudhury_mental_2014,triggs_context_2019}. Moreover, several participants were lurkers rather than active posters, although most did engage in the community through upvotes/downvotes and at times, through commenting. Overall, the personal-but-not-the-person tension highlights how interpersonal interactions in online communities like those on Reddit, even very small ones, are not necessarily about dyadic relationships but more about finding specific experiences that resonate in a community for a user. This resonance is likely to foster identity-based attachments \cite{ren_applying_2007,ren_building_2012} to the group at large, which in turn might generate a sense of belonging---a major factor in motivating participation \cite{lampe_motivations_2010}.

\subsection{Nested niches and intercommunity relationships}
\label{section:nested}
Our participants naturally described the communities they participated in \emph{in relation} to the internet, Reddit as a whole, or other subreddits. Often, subreddits are part of networks of related subreddits, with subreddits frequently explicitly connected to one another by links or adjacency of topic matter. 

Both as offshoots to larger communities and as parts of curated collections of content by users, small communities can be seen as niches-within-niches, or \textit{nested niches} as we noted in \S\ref{section:curating_online}. As subspaces to broader topical niches, the nested nature of the small communities highlights how online communities do not exist in isolation but instead are ``embedded'' \citep{uzzi_sources_1996} in an interrelated informational ecosystem.
This embeddedness appears to provide important advantages to both large and small communities, and suggests that they act in a sort of symbiosis. Indeed, we suggest that small communities may be critical for maintaining engagement on platforms like Reddit as a whole, for a few reasons. First, when each community is focused on its own ``singularity of purpose'', users are able to more effectively curate their experience and identify the kind of content that they want for a given topic. This allows large communities to avoid, for example, repetitive content from newbies by directing them to a beginner-focused subreddit.\footnote{For example, the community r/python has a sister community, r/learnpython for newcomers.} Second, these networks of related communities serve as a source of new users and new content for each other, while still allowing users the control over which subset of communities they are exposed to. Third, as the resulting content is more aligned with what users want, the informational partitioning offered by small communities may potentially foster greater loyalty to the platform for users, following the finding in \citet{tausczik_building_2014} that attachment to subgroups can foster greater loyalty to the larger group. Said differently, the smaller and larger communities together help meet the diverse needs and motivations that participants have \cite{lampe_motivations_2010}.

Similarly, as part of a platform, communities which would not have enough ``critical mass'' to survive on their own are likely sustained by the larger community. Because small communities exist within this ecosystem as nested niches, the boundaries between a given subreddit and Reddit in general can often be blurred. In this way, the concept of nested niches also underscores the role of platform affordances. Reddit's ``front page'' as well as links to related communities on subreddit sidebars and the use of cross-posting, allows for a co-mingling of content from different communities that can make connections between communities more salient. On the other hand, participants repeatedly complained that smaller communities are hurt by the Reddit front page algorithm, which prioritizes larger and more active subreddits. Our results suggest that these community dynamics may therefore differ between platforms that aggregate content (like Reddit or Facebook Groups) and those where group boundaries are more distinct (like Telegram or Discord).

Our findings underscore the importance of studying and theorizing about the interdependence of online communities, a turn which is already underway. For example, \citet{tan_tracing_2018} offers one way of exploring connections between communities, by tracing new communities to their `parent' communities through shared membership; \citet{teblunthuis_community_2021} examine dynamics of competition between communities that share the same topic space; and \citet{zhu_selecting_2014} show how community outcomes can be influenced by topic and membership overlaps.

\section{Limitations and future work}
While we cover a range of participants and subreddits, this study is limited in a number of ways. 
An important limitation is that our interview set does not provide a comprehensive picture of small online communities, a caveat we noted at the beginning of this work. For example, we do not speak to participants from small communities that have failed, died off, or grown much larger. Because we are interested in why people stay in and value small communities, we focused on persistently small communities, which are more likely to reflect rationales for why people participate in small communities (versus a more comprehensive sampling strategy).
It is likely that there are other types of small communities which are more short-lived or which go in and out of activity, which may present yet other motivations for participation.

We used a script to randomly generate a list of communities fitting our size criteria to reach out to. While reaching out to communities, however, communities that appeared more anti-social (name-calling and swearing) either did not respond or did not wish to have a recruitment post. NSFW communities did appear in our list of communities but we reached recruitment limits before moving down the list to them. This limits the types of interactions and experiences participants recounted, though we did hear tangentially about anti-social and NSFW communities through conversations with participants, which helped place participants' experiences in context. While further work should indeed investigate these different types of communities, the different kinds of online communities are so numerous that it is unlikely that any single study can encapsulate them all. We believe our data is still able to capture a range of experiences in a varied set of communities that highlights ways that participants value small online communities and evidence for how such communities can thrive and survive for years.

Finally, our study was focused specifically on the particular platform of Reddit, and it is worth considering whether our findings would generalize to other platforms. For example, Reddit has a number of affordances (e.g., an algorithmically ordered front page, pseudo-anymous user names, and subreddit sidebars) and user demographics---overwhelmingly English-speaking and from the Global North---that may distinguish it from other online community platforms. However, we believe that most of our findings are independent of the specifics of Reddit and will apply across most online communities. The motivations our participants cited, such as partitioning the information landscape or controlling who they interact with are likely to be common in any set of online communities. In addition, many of Reddit's core features (such as its "front page" algorithm and content aggregation) are common across many contemporary online community platforms. 
Regardless, future work should look at motivations for participating in small communities on diverse platforms with different affordances. In particular, it is worth exploring how well our findings translate to platforms where communities are kept more separate (as noted in \S\ref{section:nested}) or where users are not pseudo-anonymous.

Overall, future work should take small communities more seriously. By examining communities of diverse sizes and configurations, we argue that scholars can provide deeper insight into the dynamics among communities that allow them to collectively succeed on multi-community platforms. In particular, the concept of nested niches suggests unique benefits that small communities can provide to both users and platforms. 
First, researchers should further investigate different kinds of small online communities that were tangentially mentioned in interviews, such as communities that have died off and communities that are spaces for `bad behaviors' as well as mid-sized or very small communities. Related quantitative work should explore the dynamics and behavior of members of small communities. Researchers should also examine the dynamics within and across sets of interrelated communities \cite[e.g.,][]{teblunthuis_community_2021,zhu_impact_2014}, and particularly how users experience relationships between similar communities and decide how to allocate their participation. Finally, looking at small online communities that are on other sites and platforms would help us better understand to what extent these findings are universally true or a result of platform affordances.

\section{Conclusion}
Theories of online community participation suggest that communities will either grow large or die. However, online communities can actually remain quite small and most do. Through our interview study, we show that small communities are experienced in a qualitatively different way than large communities, holding a distinct role for participants. We find that participants feel their needs are met in these small communities, which provide uniquely specific informational and interactional spaces for them; that small community participants seek information and expertise in these communities rather than dyadic, interpersonal bonds, contrary to predictions from previous research. These dynamics are made possible by the fact that small communities exist in a complex ecosystem of Reddit communities and are part of a set of partially overlapping communities an individual engages with. The fact that small communities exist in tandem with larger ones point to the need to diversify our notions of community success: rather than evaluating success of a community by the metrics or relationships within it alone, success must be contextualized by the place of the community in a broader system. 
More broadly, we suggest that more explicit attention to small online communities will provide insights into how diverse types of communities collectively thrive on social computing platforms.

\begin{acks}
The authors would like to thank the interview participants for their generosity in sharing their time and experiences on these online communities, and reviewers for their thoughtful feedback in improving this manuscript. We also thank members of the Community Data Science Collective who provided multiple rounds of feedback on this project.
Special thanks also to undergraduate research assistant Daryn McElroy, who assisted in cleaning and coding interviews.
Support for this work came from the National Science Foundation (IIS-1910202, GRFP-2020303196), Northwestern University, and Purdue University.
\end{acks}

\bibliographystyle{ACM-Reference-Format}
\bibliography{niche-ref_noNotes}

\appendix

\section{Research Methods}

\subsection{Non-representative sampling for interviews}

\begin{table}[h]
\caption{A statistically non-representative stratified sampling table \cite{trost_statistically_1986} used to guide recruitment for our qualitative interview study. U.C. stands for unique contributors in a month.}

\centering
\begin{tabular}{|c|c|c|c|c|c|} 
\hline
\multicolumn{6}{|c|}{\textbf{Degree of smallness}}                                                          \\ 
\hline
\multicolumn{2}{|c|}{1-10 U.C.} & \multicolumn{2}{c|}{10-100 U.C.} & \multicolumn{2}{c|}{100+ U.C.}  \\ 
\hline
Hobbyist & Non-hobbyist        & Hobbyist & Non-hobbyist          & Hobbyist & Non-hobbyist \\
\hline
\end{tabular}
\end{table}

\received{April 2021}
\received[revised]{July 2021}
\received[accepted]{July 2021.}

\end{document}